\documentclass[prb,twocolumn,showpacs]{revtex4}
\usepackage{graphicx} 

\usepackage{amsmath,amsthm,amssymb,mathrsfs}
\usepackage{bm}

\begin{document}

\title{Crossover between fast and slow excitation of magnetization by spin torque}

\author{Tomohiro Taniguchi}
 \affiliation{
 National Institute of Advanced Industrial Science and Technology (AIST), Spintronics Research Center, Tsukuba, Ibaraki 305-8568, Japan
 }

 \date{\today} 
 \begin{abstract}
  {
A crossover between two mechanisms destabilizing the magnetization in equilibrium by the spin transfer effect is found 
in a ferromagnetic multilayer consisting of an in-plane magnetized free layer and a perpendicularly magnetized pinned layer, 
where an in-plane magnetic field is applied, and electric current flows from the pinned to the free layer. 
A fast transition from the in-plane to the out-of-plane state occurs in the low-field region, 
whereas a slow transition with small-amplitude oscillation becomes dominant in the high-field region. 
On the other hand, only the fast transition mechanism appears for the opposite current direction. 
  }
 \end{abstract}

 \maketitle


Excitation of magnetization dynamics such as switching and self-oscillation 
in a ferromagnetic/nonmagnetic multilayer by the spin transfer effect \cite{slonczewski96,berger96} 
has been studied extensively for application in practical devices 
such as magnetic memory and microwave generators 
\cite{katine00,kiselev03,kent04,lee05,kubota05,houssameddine07,ebels08,zhu08,suto12,kubota13,kudo14,bosu16,hiramatsu16}. 
It has been recognized that there is a threshold value of the electric current 
necessary to destabilize the magnetization in equilibrium and excite any type of magnetization dynamics by the spin torque effect. 
Evaluation of the threshold current is important 
because it determines the performance of spin torque devices, for example, power consumption. 
To this end, a deep understanding of the physical mechanism of magnetization instability due to the spin torque is necessary. 




In this letter, the physical mechanism destabilizing the magnetization in a ferromagnetic multilayer 
consisting of an in-plane magnetized free layer and a perpendicularly magnetized pinned layer is studied theoretically. 
We find that this system shows a crossover between two mechanisms destabilizing the magnetization in equilibrium, 
depending on the magnitude of an in-plane applied magnetic field. 
A fast transition, on the order of nanoseconds, from an in-plane stable state to an out-of-plane state is dominant in the low-field region, 
whereas a slow transition in a time range exceeding 100 ns with a small-amplitude oscillation principally determines the instability threshold in the high-field region. 
This crossover appears only when the electric current flows perpendicular to the plane from the pinned layer to the free layer. 
When the current direction is reversed, the instability threshold is determined solely by the fast transition. 



\begin{figure}
\centerline{\includegraphics[width=0.7\columnwidth]{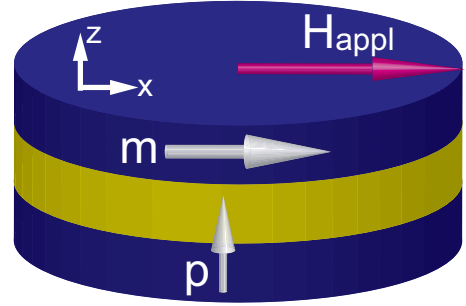}}
\caption{
         Schematic view of the system under consideration. 
         The unit vectors pointing in the magnetization directions of the free and pinned layers are denoted as 
         $\mathbf{m}$ and $\mathbf{p}$, respectively. 
         The magnetic field is applied in the in-plane direction parallel to the $x$-axis. 
         \vspace{-3ex}}
\label{fig:fig1}
\end{figure}




Figure \ref{fig:fig1} schematically shows the system under consideration. 
The $z$-axis is perpendicular to the film-plane, 
and an in-plane magnetic field is applied along the $x$-axis. 
The unit vectors pointing in the magnetization directions of the free and pinned layers are denoted by $\mathbf{m}$ and $\mathbf{p}$, respectively. 
The magnetization of the pinned layer points in the positive $z$-direction, $\mathbf{p}=+\mathbf{e}_{z}$. 
The magnetization dynamics in the free layer is described by the LLG equation, 
\begin{equation}
  \frac{d \mathbf{m}}{dt}
  =
  -\gamma
  \mathbf{m}
  \times
  \mathbf{H}
  -
  \gamma 
  H_{\rm s}
  \mathbf{m}
  \times
  \left(
    \mathbf{p}
    \times
    \mathbf{m}
  \right)
  +
  \alpha
  \mathbf{m}
  \times
  \frac{d \mathbf{m}}{dt},
  \label{eq:LLG}
\end{equation}
where $\gamma$ and $\alpha$ are the gyromagnetic ratio and Gilbert damping constant, respectively. 
The magnetic field consists of an applied field $H_{\rm appl}$ and 
demagnetization field along the $z$-direction. 
Throughout this letter, the magnetic field is considered to point in the $x$-direction. 
Then the effective magnetic field $\mathbf{H}$ in Eq. (\ref{eq:LLG}) is 
\begin{equation}
  \mathbf{H}
  =
  H_{\rm appl}
  \mathbf{e}_{x}
  -
  4\pi M 
  m_{z} 
  \mathbf{e}_{z},
  \label{eq:field}
\end{equation}
where $M$ is the saturation magnetization. 
The magnetic field is related to the magnetic energy density $E$ via $E=-M \int d \mathbf{m}\cdot \mathbf{H}$. 
In the absence of a current, the magnetization stays in the equilibrium (minimum energy) state in 
the in-plane state, $\mathbf{m}_{\rm min}=+\mathbf{e}_{x}$. 
The energy density is classified into in-plane and out-of-plane regions, 
which are divided by the saddle point $\mathbf{m}_{\rm d}=-\mathbf{e}_{x}$ \cite{taniguchi16}. 
The strength of the spin torque, $H_{\rm s}$, is given by 
\begin{equation}
  H_{\rm s}
  =
  \frac{\hbar \eta j}{2e (1+\lambda \mathbf{m}\cdot\mathbf{p}) Md},
  \label{eq:H_s}
\end{equation}
where $\eta$ and $\lambda$ are the spin polarization and spin torque asymmetry, respectively. 
The current density is denoted as $j$, 
whereas $d$ is the thickness of the free layer. 
Positive current corresponds to electron flow from the free to the pinned layer. 
The values of the parameters used in the following calculations 
are taken from typical experiments \cite{hiramatsu16}: 
$M=1300$ emu/c.c., 
$\gamma=1.764 \times 10^{7}$ rad/(Oe s), 
$\alpha=0.01$, 
$d=2$ nm, 
$\eta=0.5$, and $\lambda=\eta^{2}$.


Note that the parameter $\lambda$ has often been assumed to be zero in previous works \cite{ebels08,zhu08,suto12,kudo14,sun00}, 
including our recent work \cite{taniguchi16}, for simplicity. 
However, this parameter plays a key role, as described in the following discussion. 
The relation between $\lambda$ and the material parameters was derived theoretically in both 
a current-perpendicular-to-plane giant magnetoresistive system and magnetic tunnel junctions \cite{slonczewski96,slonczewski05,kovalev06}. 
The parameter $\lambda$ is usually positive. 




\begin{figure}
\centerline{\includegraphics[width=1.0\columnwidth]{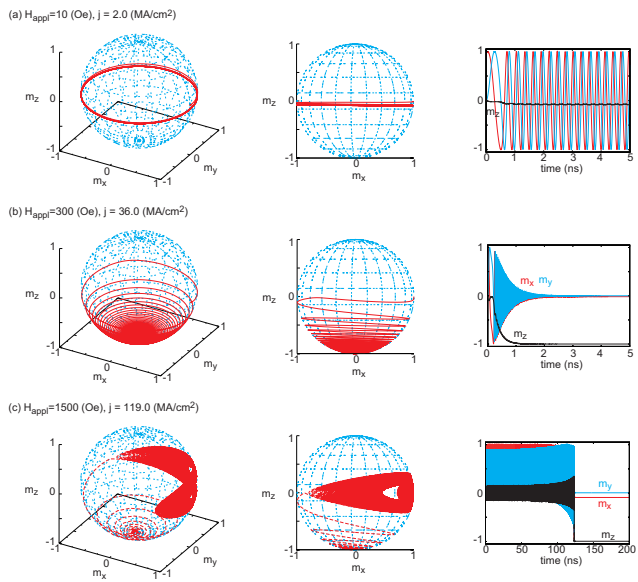}}
\caption{
         Magnetization dynamics excited by a positive current close to the threshold values. 
         The values of the in-plane applied magnetic field and current density are 
         (a) ($10$ Oe, $2.0 \times 10^{6}$ A/cm${}^{2}$), 
         (b) ($300$ Oe, $36 \times 10^{6}$ A/cm${}^{2}$), 
         and (c) ($1500$ Oe, $119 \times 10^{6}$ A/cm${}^{2}$). 
         Note that the time ranges of (a) and (b) are 5 ns, whereas that of (c) is 200 ns. 
         \vspace{-3ex}}
\label{fig:fig2}
\end{figure}



Let us show examples of magnetization dynamics obtained by solving Eq. (\ref{eq:LLG}) numerically. 
Figure \ref{fig:fig2}(a) shows the trajectory of the magnetization dynamics 
and the time evolution of each components of $\mathbf{m}=(m_{x},m_{y},m_{z})$ 
for $H_{\rm appl}=10$ Oe. 
The current density is $2.0 \times 10^{6}$ A/cm${}^{2}$, which is slightly above the threshold value. 
Starting from the stable state, 
the magnetization immediately shows a large amplitude out-of-plane self-oscillation. 
On the other hand, when the field magnitude is $300$ Oe, 
the magnetization ultimately converges to $\mathbf{m} \simeq -\mathbf{e}_{z}$ 
and its dynamics stops without showing stable self-oscillation, 
as shown in Fig. \ref{fig:fig2}(b), where the current density is $36 \times 10^{6}$ A/cm${}^{2}$. 
The absence of the self-oscillation indicates that the threshold current density is larger than the current 
necessary to stabilize the self-oscillation. 
This dynamics is consistent with our previous work \cite{taniguchi16}. 


It is newly reported in this letter that 
the instability mechanism undergoes a transition when the field magnitude is further increased. 
An example of this dynamics is shown in Fig. \ref{fig:fig2}(c), 
where $H_{\rm appl}=1500$ Oe and $j=119 \times 10^{6}$ A/cm${}^{2}$. 
The magnetization shows an in-plane precession before it moves to the point $\mathbf{m} \simeq -\mathbf{e}_{z}$. 
We emphasize that a relatively long period of time is necessary to move to the out-of-plane state, compared with the low-field case. 
For example, it takes more than 100 ns for the high-field case shown in Fig. \ref{fig:fig2}(c), 
whereas almost 1 ns is sufficient for the low field case in Fig. \ref{fig:fig2}(b). 




\begin{figure}
\centerline{\includegraphics[width=1.0\columnwidth]{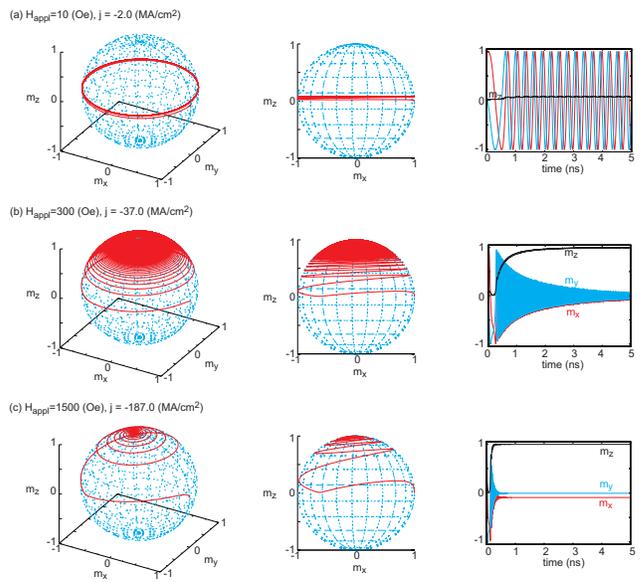}}
\caption{
         Magnetization dynamics excited by a negative current close to the threshold values. 
         The values of the in-plane applied magnetic field and current density are 
         (a) ($10$ Oe, $-2.0 \times 10^{6}$ A/cm${}^{2}$), 
         (b) ($300$ Oe, $-37 \times 10^{6}$ A/cm${}^{2}$), 
         and (c) ($1500$ Oe, $-187 \times 10^{6}$ A/cm${}^{2}$). 
         \vspace{-3ex}}
\label{fig:fig3}
\end{figure}



We find that a negative current can also destabilize the magnetization, 
but a crossover between different mechanisms of instability does not exist. 
Figures \ref{fig:fig3}(a)-(c) show the trajectories and time evolutions of the magnetization dynamics, 
where the field magnitudes are the same as those shown in Figs. \ref{fig:fig2}(a)-(c), respectively. 
The values of the current density are close to the threshold values,
(a) $-2.0$, (b) $-37$, and (c) $-187$ $\times 10^{6}$ A/cm${}^{2}$. 
In all cases, the magnetization immediately moves to the out-of-plane state in a short period of time without showing an in-plane precession, 
and shows large amplitude self-oscillation or arrives at the point $\mathbf{m} \simeq +\mathbf{e}_{z}$, 
depending on the field magnitude. 

\begin{figure}
\centerline{\includegraphics[width=1.0\columnwidth]{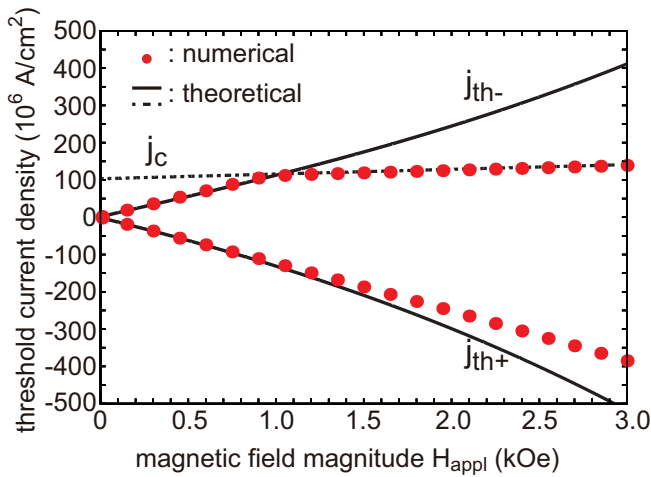}}
\caption{
         Dependence of the threshold current density evaluated by solving Eq. (\ref{eq:LLG}) numerically (dots). 
         Equation (\ref{eq:jc}) and (\ref{eq:j_th}) are also indicated by dotted and solid lines, respectively. 
         \vspace{-3ex}}
\label{fig:fig4}
\end{figure}

The dots in Fig. \ref{fig:fig4} summarize the dependence of the threshold current density 
on the in-plane applied field magnitude \cite{comment}. 
In both positive and negative current regions, 
the threshold current density increases with increasing field magnitude. 
However, the field dependence of the threshold current changes at $H_{\rm appl} \simeq 1$ kOe in the positive current region, 
where the instability mechanism changes from the fast transition shown in Figs. \ref{fig:fig2}(a) and \ref{fig:fig2}(b) 
to the slow one in Fig. \ref{fig:fig2}(c). 
Figure \ref{fig:fig4} is the central result in this letter. 



The above numerical results are explained by two theories focusing on different instability mechanisms. 
One approach, which has been widely used to analyze the spin-torque-induced magnetization dynamics, 
is called linear analysis of the LLG equation \cite{ebels08,sun00,grollier03}. 
In this approach, the spin torque is assumed to cancel the damping torque 
when the current magnitude equals that of the threshold current. 
Therefore, the magnetization shows a small-amplitude oscillation around a stationary state 
because of the precessional torque, 
which is given by the first term on the right hand side of Eq. (\ref{eq:LLG}). 
When the current magnitude is increased slightly, the spin torque slightly overcomes the damping torque, 
and the magnetization moves from the stable state slowly, accompanied by precession. 
In the present system, the stable state is $\mathbf{m}=+\mathbf{e}_{x}$, 
and the linearized LLG equation describes the small amplitude oscillations of $m_{y}$ and $m_{z}$ as 
\begin{equation}
  \frac{1}{\gamma}
  \frac{d}{dt}
  \begin{pmatrix}
    m_{z} \\
    m_{y} 
  \end{pmatrix}
  +
  \mathsf{M}
  \begin{pmatrix}
    m_{z} \\
    m_{y}
  \end{pmatrix}
  =
  \frac{\hbar \eta j}{2eMd}
  \begin{pmatrix}
    1 \\
    0 
  \end{pmatrix},
  \label{eq:LLG_linear}
\end{equation}
where the components of the $2 \times 2$ matrix $\mathsf{M}$ are 
$\mathsf{M}_{1,1} = \alpha (H_{\rm appl}+4 \pi M)-[\hbar \eta \lambda j/(2eMd)]$, 
$\mathsf{M}_{1,2} = H_{\rm appl}$,
$\mathsf{M}_{2,1} = -H_{\rm appl} - 4 \pi M$, 
and $\mathsf{M}_{2,2} = \alpha H_{\rm appl}$. 
The solution of Eq. (\ref{eq:LLG_linear}) is described by 
$\exp\{\gamma[ \pm i \sqrt{{\rm det}[\mathsf{M}]-({\rm Tr}[\mathsf{M}/2])^{2}} - {\rm Tr}[\mathsf{M}]/2]t\}$. 
The threshold current of the linearized LLG equation is defined as the current 
satisfying ${\rm Tr}[\mathsf{M}]=0$ \cite{grollier03} 
and is given by 
\begin{equation}
  j_{\rm c}
  =
  \frac{4 \alpha eMd}{\hbar \eta \lambda}
  \left(
    H_{\rm appl}
    +
    2\pi M
  \right).
  \label{eq:jc}
\end{equation}


We emphasize that Eq. (\ref{eq:jc}) has a fixed sign (positive in the present case), 
and diverges in the limit of $\lambda \to 0$. 
This is because only the spin torque excited by the positive current has a component 
antiparallel to the damping torque 
after the spin torque is averaged over the precession when $\lambda>0$, 
and therefore, can cancel the damping torque. 

Another approach to deriving the theoretical formula for the threshold current 
focuses on energy absorption of the magnetization from the work done by the spin torque 
during a time shorter than the precession period \cite{taniguchi16}. 
In this case, the threshold current is defined as the current satisfying 
\begin{equation}
  \int_{\mathbf{m}_{\rm min}}^{\mathbf{m}_{\rm d}}
  dt 
  \frac{dE}{dt}
  =
  E_{\rm saddle}
  -
  E_{\rm min}.
  \label{eq:dEdt}
\end{equation}
According to Eq. (\ref{eq:dEdt}), the magnetization absorbs energy 
sufficiently larger than the energy barrier, $E_{\rm saddle}-E_{\rm min}$, 
between the in-plane stable state and the out-of-plane state from the work done by the spin torque. 
The integral path in Eq. (\ref{eq:dEdt}) moves directly from the minimum energy state to the saddle point without any precession, 
indicating that Eq. (\ref{eq:dEdt}) can be used to evaluate the transition during a time shorter than the precession period. 
When the applied field magnitude $H_{\rm appl}$ is much smaller than the demagnetization field, $4\pi M$, 
the threshold current density satisfying Eq. (\ref{eq:dEdt}) is approximately given by 
\begin{equation}
  j_{\rm th \pm}
  =
  \mp
  \frac{2 eMd}{\hbar \eta}
  4\pi M 
  \frac{\mathcal{N}}{\mathcal{D}_{\pm}},
  \label{eq:j_th}
\end{equation}
where $\mathcal{N}$ and $\mathcal{D}_{\pm}$ are 
\begin{equation}
\begin{split}
  \mathcal{N}
  =&
  4 \lambda^{2}
  \left[
    2 \alpha 
    \left(
      3 
      -
      2 h 
    \right)
    \left(
      1
      -
      h
    \right)
    \sqrt{
      h(1-h)
    }
    +
    3h
  \right]
\\
  & 
  \times
  \sqrt{1-4\lambda^{2} h(1-h)},
\end{split}
\end{equation}
\begin{equation}
\begin{split}
  \mathcal{D}_{\pm}
  =&
  3 
  \left\{
    \sqrt{
      1
      -
      4 \lambda^{2}
      h (1-h)
    }
    \left[
      \pi
      \mp
      4 \lambda
      \sqrt{h(1-h)}
    \right]
  \right.
\\
  &
  \left.
    -
    2 \left[
      1
      -
      2 \lambda^{2}
      \left(
        1
        -
        h
      \right)
    \right]
    \cos^{-1}
    \left[
      \pm 2 \lambda
      \sqrt{h(1-h)}
    \right]
  \right\},
\end{split}
\end{equation}
and $h=H_{\rm appl}/(4\pi M)$. 
The current $j_{\rm th+(-)}$ is the current necessary to move the magnetization to the positive (negative) $z$ region, 
and its sign is negative (positive). 
In this approach, both positive and negative current can destabilize the magnetization. 
This is because, if we focus on the time shorter than the precession period, 
the spin torque can have component antiparallel to the damping torque for both current directions. 


In Fig. \ref{fig:fig4}, we also show Eqs. (\ref{eq:jc}) and (\ref{eq:j_th}) by dotted and solid lines, respectively. 
In the positive current region, the numerically evaluated threshold current density, shown by dots, is well fitted by ${\rm min}[j_{\rm c},j_{\rm th-}]$. 
For a low field ($H_{\rm appl} \lesssim 1$ kOe for the present parameters), the threshold current density is given by $j_{\rm th-}$, from Eq. (\ref{eq:j_th}), 
which is derived by focusing on the fast transition from the in-plane to the out-of-plane state during a time shorter than the precession period. 
In fact, as shown in Figs. \ref{fig:fig2}(a) and \ref{fig:fig2}(b), the transition occurs within a relatively short time. 
On the other hand, for a high field, the threshold current is given by $j_{\rm c}$, from Eq. (\ref{eq:jc}), 
which is derived from the linearized LLG equation, which assumes small-amplitude oscillation around the stationary state. 
This is also consistent with the numerical result shown in Fig. \ref{fig:fig2}(c), where the in-plane precession was observed before the transition. 
On the other hand, the threshold current density in the negative current region is well fitted by $j_{\rm th+}$. 
This is consistent with the numerical result in Fig. \ref{fig:fig3} where the instability occurs in a short period of time 
in both low- and high-field regions. 
The difference between the numerical and theoretical results for negative current in the relatively high-field region is due to the fact that 
Eq. (\ref{eq:j_th}) is valid for $h=H_{\rm appl}/(4\pi M) \ll 1$. 


We emphasize that the spin torque asymmetry $\lambda$ plays a key role in observing the crossover of 
the two physical mechanisms found in the positive current region in Fig. \ref{fig:fig4}. 
When $\lambda$ is zero, Eq. (\ref{eq:jc}) diverges; 
thus, ${\rm min}[j_{\rm c},j_{\rm th-}]$ is always $j_{\rm th-}$. 
This means that only the fast transition from the in-plane to the out-of-plane state occurs, 
similar to the situation in the negative current region; 
therefore, a crossover does not exist \cite{taniguchi16}. 
On the other hand, for finite $\lambda$, the magnitude relation between $j_{\rm c}$ and $j_{\rm th-}$ changes 
at a certain field magnitude ($H_{\rm appl} \simeq 1$ kOe for the present parameters); 
thus, the crossover is observed. 


The values of the parameters used in the above calculations are derived from typical experiments 
\cite{houssameddine07,suto12,bosu16,hiramatsu16}. 
As shown in Fig. \ref{fig:fig4}, the current density required to observe the crossover ($\sim 10^{8}$ A/cm${}^{2}$) is relatively high. 
Therefore, a giant magnetoresistive system, rather than a magnetic tunnel junction, will be preferable 
for observing the crossover experimentally. 


In conclusion, we evaluated the threshold current required to destabilize the magnetization in equilibrium by spin torque 
in a ferromagnetic multilayer consisting of an in-plane magnetized free layer 
and a perpendicularly magnetized pinned layer. 
The central result is shown in Fig. \ref{fig:fig4}, 
where a crossover between two physical mechanisms destabilizing the magnetization is observed in the positive current region. 
A fast transition from the in-plane stable state to the out-of-plane state occurs in the low-field region, 
whereas a slow transition with the small-amplitude oscillation around the stable state is dominant in the high-field region. 
On the other hand, only the fast transition occurs in the opposite current direction. 


The author is grateful to Takehiko Yorozu, Hitoshi Kubota, Ryo Hiramatsu, Sumito Tsunegi, and Shingo Tamaru 
for valuable discussions. 
The author is also thankful to Satoshi Iba, Yoichi Shiota, Aurelie Spiesser, Hiroki Maehara, and Ai Emura 
for their support and encouragement. 
This work is supported by the Japan Society and Technology Agency (JST) strategic innovation promotion program 
"Development of new technologies for 3D magnetic recording architecture". 



\begin{thebibliography}{21}
\bibitem[{com()}]{comment}
\bibinfo{note}{The threshold current in the numerical simulation is the minimum
  current satisfying $|m_{x}(t=1 \mu{\rm s})-m_{x}(t=1\mu{\rm s}-\Delta
  t)|>10^{-10}$ or $|m_{z}(t=1\mu{\rm s})|>0.9$, where $\Delta t=50$fs is the
  time step of the simulation. The former condition means that the
  magnetization is in the self-oscillation state, while the latter means that
  the magnetization moves to the out-of-plane state.}
\end{thebibliography}


\end{document}